\documentclass[twocolumn,PRB,amsmath,amssymb]{revtex4-1}
\usepackage{amssymb}
\usepackage{amsmath}
\usepackage{graphicx}
\usepackage{epstopdf}
\usepackage{bm}%
\usepackage{gensymb}
\usepackage{soul}
\usepackage{sidecap}
\usepackage[normalem]{ulem}
\usepackage[colorlinks,citecolor=blue,linkcolor=blue,anchorcolor=blue,filecolor=blue,urlcolor=blue]{hyperref}
\usepackage{epsfig}
\usepackage{graphicx}
\usepackage{amsfonts,amssymb}
\usepackage{amsmath}
\usepackage{color}
\usepackage {times}
\usepackage{epstopdf}

\newcommand{\bl}{\begin{aligned}}
\newcommand{\el}{\end{aligned}}

\newcommand{\be}{\begin{equation}}
\newcommand{\ee}{\end{equation}}   
\newcommand{\bea}{\begin{eqnarray}}
\newcommand{\eea}{\end{eqnarray}}
\newcommand{\ba}{\begin{array}}
\newcommand{\ea}{\end{array}}

\newcommand{\q}{{\bf q}}
\renewcommand{\k}{{\bf k}}
\newcommand{\Q}{{\bf Q}}

\graphicspath{{Figs/}}

\begin{document}

\title{Role of Dirac cones in the anisotropic properties associated with the spin-density wave state of iron pnictides}

\date{\today}
\author{Garima Goyal}
\author{Dheeraj Kumar Singh}
\email{dheeraj.kumar@thapar.edu }
\affiliation{Department of Physics and Material Science, Thapar Institute of Engineering and Technology, Patiala-147004, Punjab, India}

\begin{abstract}
The origin of unusual anisotropic electronic properties in the spin-density wave state of iron pnictides has conventionally been attributed to the breaking of four-fold rotational symmetry associated with the collinear magnetic order. By using a minimal two-orbital model, we show that a significant portion of the contribution to the anisotropy may come from the Dirac cones, which are not far away from the Fermi level. We demonstrate this phenomenon by examining optical conductivity and quasiparticle interference in the Dirac-semimetallic state with spin-density wave order, and the latter can be obtained by choosing appropriate interaction parameters and orbital splitting between the $d_{xz}$ and $d_{yz}$ orbitals. We further extend this study to investigate the low-energy spin-wave excitations in the Dirac-semimetallic state with spin-density wave order.
\end{abstract}

\maketitle
\newpage
\section{Introduction} 
Iron-based superconductors (IBS)~\cite{kamihara, chen1, guo, johnston, takahashi, paglione, mizuguchi, hsu, rotter, sasmal, mazin} is the largest family of material systems after high $T_c$ cuprates~\cite{bednorz, sheng}, which exhibits unconventional superconductivity~\cite{nakai, kar}. In their temperature-vs-doping phase diagram, the superconducting phase is surrounded by the nematic and spin-density wave (SDW) phases~\cite{ni, ren, basov}. Thus, all three phases appear to compete with each other, the SDW and superconducting states in particular~\cite{chu1, luetkens, zhao, drew}. On the contrary, the proximity of the superconducting state to the SDW state has often been taken as a sign of the existence of reminiscent magnetic fluctuations that are considered instrumental in binding the Cooper pairs~\cite{chubukov, chen, fang, mazin1, singh}.

The lattice distortion associated with the orthorhombic/nematic phase is found inadequate in explaining the degree of anisotropy exhibited by the electronic properties, therefore, the latter is believed to be of electronic origin~\cite{nandi, fang1, kitagawa, kitagawa1, matan, dusza}. Several earlier works have indicated that the phenomenon of nematicity is not restricted to IBS alone~\cite{nakata, xie}. For the IBS, however, it has often been characterized by a non-vanishing energy splitting between $d_{xz}$ and $d_{yz}$ orbitals~\cite{ yi, zhang, zhang1, pfau, lee}.

Anisotropic electronic properties in the SDW state with ordering wavevector ($\pi, 0$) or in the orthorhombic/nematic state are not abnormal as the four-fold rotational symmetry is already broken~\cite{chu, fischer, tanatar, valenzuela, matusiak, bascones}. However, evidence of an anisotropic electronic state extends even to the high-temperature tetragonal phase~\cite{blomberg, rosenthal} and superconducting state~\cite{song}.
Their signature has largely been obtained with the help of experiments such as angle-resolved photoemission spectroscopy (ARPES)~\cite{yi, richard}, transport measurements~\cite{chu, tanatar}, scanning tunneling microscopy (STM)~\cite{chuang} etc.
The origin of anisotropy existing across different phases has been studied extensively.

In the SDW state, the emergence of anisotropic transport properties has conventionally been attributed to the orbital-weight redistribution along the reconstructed Fermi surface~\cite{pinaki}. It can result into anisotropic impurity scattering~\cite{knolle, dheeraj}, which is also reflected in the quasiparticle interference (QPI) obtained through the STM measurements~\cite{allan, chuang, zhou}. The QPI patterns consist of a quasi-one-dimensional structure with a length scale of the order $\sim 6a - 8a$, where $a$ is the lattice constant. Various theoretical studies attribute the anisotropic patterns in the SDW state to the reconstructed elliptical hole pocket around $(0, 0)$~\cite{knolle, dheeraj, plonka} while others to a large orbital splitting between the $d_{xz}$ and $d_{yz}$ orbitals~\cite{dheeraj1}. In the latter case, the strength of the orbital-splitting parameter used to explain the anisotropic behavior cannot alone be accounted for by the four-fold rotational symmetry breaking associated with ($\pi, 0$)-type collinear order. Moreover, the anisotropic feature is not limited to the QPI patterns and transport properties, where a major contribution comes from the electronic states in the vicinity of the Fermi level, the optical spectra also show the anisotropy existing up to photonic energies $\sim 2$eV~\cite{nakijama}.

Despite significant progress made in theoretical as well as experimental studies, the origin of highly-anisotropic electronic properties of the SDW state continues to be a long-standing problem. More specifically, almost all the previous studies, for a realistic set of interaction parameters, were unable to reproduce the one-dimensional
characteristics of the QPI patterns, thereby, indicating the  limitations of the models used~\cite{knolle,plonka}. This was mainly because of the dominant feature in the QPI pattern arising due to the intrapocket scattering associated with a large Fermi pocket around the $(0, 0)$ point. It may be noted that ARPES measurements appear to negate the presence of such a large pocket around $(0, 0)$ and find only very small circle-like Fermi surfaces separated by $\sim (\pi/4,  0$) and perhaps associated with the Dirac cones~\cite{richard, watson}. This leads to the question whether these small circle-like Fermi surfaces can give rise to nearly one-dimensional nature of the QPI patterns.

The anisotropic behavior of the conductivity observed experimentally~\cite{chuang,kobayashi} was captured by theoretical studies~\cite{sugimoto, zhang2} though the origin of its unconventional nature was attributed obscurely to the interplay of correlation effects and bandstructure ~\cite{blomberg1}. If the small circle-like structures in the Fermi surfaces are to be associated with Dirac cones~\cite{richard, hasan}, then it becomes crucial to understand their role in conductivity anisotropy. The orbital-weight distribution along the Dirac cone, especially in the vicinity of the Dirac point, may be highly anisotropic as compared to the portions of other bands which may cross the Fermi level.

Theoretically, the existence of Dirac cones in the SDW state was predicted earlier~\cite{ran}. According to the bandstructure details obtained from the first-principle calculation, there exists a good nesting between the hole pocket around $(0, 0)$ and the electron pocket around ($\pi, 0$)~\cite{raghu, graser, kuroki, ikeda, brydon}. Consequently, the Fermi surface instability is expected to introduce a SDW gap all along the Fermi pockets. However, arguments based on symmetry and band topology prohibit full opening of the gap all along the Fermi pockets~\cite{ran}. This has been attributed to the vorticity mismatch of the electron and hole pockets, where the vorticity is associated with a spinor-like state vector defined by replacing the up and down spin electronic states with  $d_{xz}$ and $d_{yz}$ orbital states. Moreover, the reconstructed bands in the SDW state are accompanied by Dirac cones, not far away from the Fermi level~\cite{ran, richard, garima}, protected by a set of symmetries, which includes collinear magnetic order, inversion symmetry, and a combined symmetry operation involving simultaneous magnetic moment inversion and time reversal.

Through this paper, we clarify the role of Dirac cones, which was ignored in the earlier works, in making a significant contribution to the anisotropic electronic properties in the SDW state of iron pnictides. We achieve this objective by examining the Drude weight, optical conductivity, quasiparticle interference, etc. in the Dirac-semimetallic state within a minimal two-orbital model. An important advantage of the Dirac semimetallic state, which is easier to realize in the two-orbital model, is that there are no other bands crossing the Fermi level, therefore, the contribution of Dirac cones in causing anisotropies in various electronic properties can distinctly be demarcated. Our finding suggests that as the chemical potential approaches the Dirac point, the anisotropy maximizes. Furthermore, the interpocket scattering between the pockets associated with two distinct Dirac cones leads to a nearly one-dimensional quasiparticle pattern similar to the one observed in the SDW state. All these findings suggest an important role of the Dirac cones in the anisotropic electronic properties observed in the SDW state of iron pnictides.

\section{Model and method}
\subsection{Multi-orbital models of iron pnictides}
We consider the minimal two-orbital model proposed earlier by Raghu \textit{et al.}~\cite{raghu}, which captures the essential features of the Fermi surfaces as suggested by the bandstructure calculations. The model incorporates only the two orbitals $d_{xz}$ and $d_{yz}$. The tight-binding part of the Hamiltonian is given by
 \begin{equation}
 \mathcal{H}_k  =  \sum_{\bf{k}} \sum_{\alpha ,\beta} \sum_{\sigma} \varepsilon_{\bf{k}}^{\alpha \beta} \textit{d}_{\bf{k} \alpha \sigma}^{\dagger}  \textit{d}_{\bf{k} \beta \sigma} - \delta (\textit{d}_{\bf{k} \alpha \sigma}^{\dagger}  \textit{d}_{\bf{k} \alpha \sigma} - \textit{d}_{\bf{k} \beta \sigma}^{\dagger}  \textit{d}_{\bf{k} \beta \sigma}) ,
 \end{equation}
where $\textit{d}_{\bf{k}\alpha\sigma}^{\dagger} (\textit{d}_{\bf{k}\alpha\sigma})$
is the creation (destruction) operator for an electron with spin $\sigma$ and momentum ${\bf k}$ in the orbital $\alpha$. $\alpha/\beta$ denotes the $d_{xz}/d_{yz}$ orbitals of the iron atom. The parameter $\delta$ accounts for the orbital splitting between the two orbitals. The matrix elements $\varepsilon_{\bf{k}}^{\alpha \beta}$ are momentum-dependent orbital energies and are given by
\begin{gather}
\varepsilon^{\alpha \alpha/ \beta \beta}_x = -2t_{1/2} \cos k_x \nonumber \\ \varepsilon^{\alpha \alpha/ \beta \beta}_y = -2t_{2/1} \cos k_y \nonumber\\
\varepsilon^{\alpha \alpha/ \beta \beta}_{xy} = -4t_3 \cos k_x \cos k_y\nonumber\\
\varepsilon^{\alpha \beta}_{xy} = -4t_4 \sin k_x \sin k_y.
\end{gather}
$t_1$ and $t_2$ represent the nearest-neighbor hopping parameters along $x$ and $y$, respectively. $t_3$ and $t_4$ denote the next-nearest neighbors connecting the same and different orbitals, respectively. $t_1 = -1$, $t_2 = 1.3$, $t_3 = t_4 = -0.85$. The unit of energy is set to be in terms of $|t_1|$.

The interaction part of the Hamiltonian includes the standard on site Coulomb repulsion terms given by
 \begin{eqnarray}
 \mathcal{H}_{int}  =  \textit{U} \sum_{{\bf i}  \alpha} \textit{n}_{\bf{i} \alpha \uparrow} \textit{n}_{\bf{i}  \alpha \downarrow}  +  (\textit{U}^{\prime}  -  \frac{\textit{J}}{2}) \sum_{\bf{i}, \alpha < \beta} \textit{n}_{\bf{i} \alpha} \textit{n}_{\bf{i} \beta}  - \nonumber \\
 2 \textit{J} \sum_{\bf{i}, \alpha < \beta} \textit{\bf S}_{{\bf i} \alpha} \cdot \textit{\bf S}_{{\bf i} \beta} +  \textit{J} \sum_{\bf{i}  ,  \alpha < \beta  ,  \sigma} \textit{d}^{\dagger}_{{\bf i}\alpha \sigma} \textit{d}^{\dagger}_{{\bf i}\alpha \bar{\sigma}}\textit{d}_{{\bf i}\beta \bar{\sigma}} \textit{d}_{{\bf i}\beta \sigma}.
 \end{eqnarray}
The first and second terms describe intra and interorbital Coulomb interactions, respectively, where $\hat{n}_{{\bf i}\alpha \sigma} = d^{\dagger}_{{\bf i} \alpha \sigma} d_{{\bf i} \alpha \sigma}$ is the number operator for the particles with spin $\sigma$ at site {\bf i} in an orbital $\alpha$. The third and fourth terms represent Hund’s coupling and pair hopping, where $\bar{\sigma}$ denotes the spin anti-parallel to $\sigma$. The relation $\textit{U}   =  \textit{U}^{\prime}  +  2 \textit{J}$ is ensured to maintain the rotational symmetry.
 
\subsection{Mean-field Methodology}
The terms in the interaction Hamiltonian are quartic in operators which are decoupled into bilinear form via mean-field decoupling in order to study the SDW state with ordering wave vector $(\pi,0)$, which corresponds to ferromagnetic and antiferromagnetic arrangements of magnetic moments along $y$ and $x$ directions. For simplicity, we chose the direction of the self-consistently obtained magnetic moments along the $z$ direction without loss of generality because there is rotational symmetry. The meanfield Hamiltonian for the SDW state takes a matrix form~\cite{kovacic}
\begin{eqnarray}    \label{eqn:6}
\hat{\mathcal{H}}_{mf}  =  \sum_{\bf{k} \sigma} \Phi^{\dagger}_{\bf{k} \sigma}
\begin{pmatrix}
\hat{\varepsilon}_{\bf{k}}  + \hat{N} - \delta \tau_z  &  \hat{\Delta} \\
\hat{\Delta}   &    \hat{\varepsilon}_{\bf{k+Q}} + \hat{N} - \delta \tau_z
\end{pmatrix}
\Phi_{{\k} \sigma},
\end{eqnarray} 
 where the basis set is $\Phi^{\dagger}_{\bf{k} \sigma}$  =  $(\textit{d}^{\dagger}_{{\k} \alpha \sigma} \textit{d}^{\dagger}_{{\bf k} \beta \sigma} \textit{d}^{\dagger}_{\bf{k + Q} \alpha \sigma} \textit{d}^{\dagger}_{{\k+\Q} \beta \sigma})$. Each matrix element in the above Hamiltonian is itself a 2$\times$2 matrix in the orbital basis. $\tau$ is the Pauli matrix for the orbital basis. The exchange field $\hat{\Delta}$ and $\hat{\textit{N}}$, in terms of onsite interaction parameters, orbital magnetization and charge densities are given by
\begin{eqnarray*}
  2  \Delta_{\alpha}  =  \textit{U} \textit{m}_{\alpha}  +  \textit{J} \sum_{\alpha \neq \beta} \textit{m}_{\beta}   \label{eqn:7}  \nonumber  \\
  2   \textit{N}_{\alpha}  =  \textit{(5J - U)} \textit{n}_{\alpha},
  \end{eqnarray*}
\noindent where the order parameters $m_{\alpha/\beta}$ and $n_{\alpha/\beta}$ are calculated self-consistently using eigenvalues and eigenvectors of the meanfield Hamiltonian $\hat{\mathcal{H}}_{mf}$. \\

\subsection{Low-energy Excitations}

\subsubsection{\bf \textit{Quasiparticle Interference using T-matrix approximation}}
QPI, generated by the scattering off of a quasiparticle by an impurity atom, has been frequently used to obtain important insight into the electronic structure especially in the vicinity of the Fermi level. Theoretically, the QPI can be studied by calculating the modulation in Green's function arising as a result of the quasiparticle getting scattered by an impurity potential. In the current work, we consider a single impurity with $\delta$-potential such that the orbital state of the quasiparticle is preserved.

In the SDW state, the Green's function is given by~\cite{dheeraj}
\begin{equation}
     \hat{G}^{0} ({\bf k}, \omega) = [\omega  \hat{I} - \hat{\mathcal{H}}_{mf}]^{-1},\nonumber
\end{equation}
where $\hat{I}$ is a $4 \times 4$ identity matrix.

The modification in the Green's function due to the scattering by a non-magnetic impurity can be obtained within the $\hat{T}$ matrix approximation as follows
\begin{equation}
    \centering
      \delta \hat{G} ({\bf k_1, k_2}, \omega) = \hat{G}^{0} ({\bf k_1}, \omega) \hat{T} (\omega) \hat{G}^{0} ({\bf k_2}, \omega).
\end{equation}
Here, $\hat{T}(\omega)$ matrix is given by
\begin{equation}
    \centering
       \hat{T} (\omega) = (\hat{{I}} - \mathcal{\hat{G}}^{0} (\omega))^{-1} \hat{V}_{imp}.
\end{equation}
$\mathcal{\hat{G}}^{0} (\omega)$ is obtained by summing over all the momenta in the Brillouin zone as follows
\begin{equation}
    \centering
       \mathcal{\hat{G}}^{0} (\omega) = \frac{1}{N} \sum_{\k} \hat{G}^{0} ({\bf k}, \omega).
\end{equation}
The impurity potential, owing to the orbital and momentum basis, also takes a 4 $\times$ 4 matrix form
\begin{equation}
    \centering
       \hat{V}_{imp} = {{V}}_{0}
    \begin{pmatrix}
     {\rm I}_{2 \times 2}   &   {\rm I}_{2 \times 2} \\
     {\rm I}_{2 \times 2}   &   {\rm I}_{2 \times 2}
    \end{pmatrix}.
     \nonumber
\end{equation}
${\rm I}_{2 \times 2}$ is a $2 \times 2$ identity matrix and $V_o$ is the parameter denoting the strength of impurity potential.

The modification $\delta \rho_{ij} ({\bf k}, \omega)$ induced in the DOS by the impurity in the momentum space is obtained as
\begin{eqnarray}
    \centering
       \delta \rho ({\bf q}, \omega) &=& -\frac{i}{2 \pi} \sum_{\bf k} g({\k}, {\q}, \omega)
\end{eqnarray}
with
\begin{equation}
    g({\k}, {\q}, \omega) = Tr \delta \hat{G}({\k}, {\k+\q}, \omega) - Tr \delta \hat{G}({\k+\q}, {\k}, \omega). \nonumber
\end{equation}

The real space QPI pattern can be calculated via Fourier transform as follows
\begin{equation}
    \delta \rho ({\bf r}_i, \omega) = \frac{1}{N} \sum_{\bf k} \delta \rho ({\bf q}, \omega) e^{i {\bf k} \cdot {\bf r}_i}.
\end{equation}
 In the calculations, the strength of impurity potential $V_0$ is set to be 0.2 and a mesh size of 300 $\times$ 300 is considered.

\subsubsection{\bf \textit{Optical Conductivity}}
In order to study the effect of Dirac cone on the charge dynamics, we investigate the optical conductivity $\sigma_l$ along $l = x \,\,{\rm or}\,\, y$ directions with the antiferromagnetic and ferromagnetic arrangement of magnetic moments, respectively. The components of the optical conductivity are given by~\cite{dagotto, valenzuela1}
\begin{eqnarray} \sigma_l &=& D_l \delta{(\omega)}+
\frac {1}{N} \sum_{{\bf k}, n \ne n^{\prime}} 
\frac{ |j^l_{nn^{\prime}}({\bf k})|^2}{{\varepsilon}_{n^\prime {\bf k}} -
{\varepsilon}_{n {\bf k}}} \nonumber\\
&\times&\theta (-{\varepsilon}_{n^\prime {\bf k}})\theta ({\varepsilon}_{n {\bf k}})
\delta(\omega-{\varepsilon}_{n^\prime {\bf k}}+{\varepsilon}_{n {\bf k}}),
\end{eqnarray}
where $D_l$ is the Drude weight can be obtained as
\begin{eqnarray} \frac {D_l}{2 \pi} &=& \frac{1}{2N}
\sum_{\bf k} T^l_{nn} ({\bf k}) \theta (-{\varepsilon}_{n {\bf k}}) -
\frac {1}{N} \sum_{{\bf k}, n \ne n^{\prime}} 
\frac{ |j^l_{nn^{\prime}}({\bf k})|^2}{{\varepsilon}_{n^\prime {\bf k}} -
{\varepsilon}_{n {\bf k}}} \nonumber\\
&\times&\theta (-{\varepsilon}_{n^\prime {\bf k}})\theta ({\varepsilon}_{n {\bf k}}).
\end{eqnarray}
Here, ${\varepsilon}_{n{\bf k}}$ is the single particle energy in the SDW state, $\theta$ is the step function, and the energy subscript $n\, ({\rm or}\,\, n^{\prime})$ is the band index. Moreover,
\begin{eqnarray} T^{l}_{nn} = \sum_{\alpha \beta} T^{l;\alpha \beta}_{nn}
= \sum_{\alpha \beta} \frac {\partial^2 
{\varepsilon}_{\alpha \beta} ({\bf k})}{\partial k^2_l}
c^*_{{\bf k} \alpha n} c_{{\bf k}\beta n}
\nonumber\\
j^{l}_{nn^{\prime}} = -\sum_{\alpha \beta} j^{l;\alpha \beta}_{nn^{\prime}} 
= -\sum_{\alpha \beta} \frac {\partial
{\varepsilon}_{\alpha \beta} ({\bf k})}{\partial k_l}
c^*_{{\bf k} \alpha n} c_{{\bf k}\beta n},
\end{eqnarray}
where $c_{{\bf k} \alpha  n}$ is the matrix element belonging to the unitary transformation from the orbital to the band basis. The $\delta$ function is approximated by the Lorentzian, where a small broadening parameter is used in both directions of the same magnitude.
In order to gain insight into the origin of the anisotropy, we define orbital-dependent components of the Drude weight as follows
\begin{gather}
    \frac{D_l^{\alpha \beta}}{2 \pi} = \frac{1}{2N} \sum_{\k} T_{nn}^{l; \alpha \beta} ({\k}) \theta (-{\varepsilon}_{n{\bf k}})  \nonumber \\
    - \frac{1}{N} \sum_{{\bf k}, n \neq n'} Re \frac{j_{nn'}^{l; \alpha \beta^{*}} ({\bf k}) j_{nn'}^{l} ({\bf k})}{{\varepsilon}_{n' {\bf k}} - {\varepsilon}_{n {\bf k}}} \theta (-{\varepsilon}_{n' {\bf k}}) \theta ({\varepsilon}_{n {\bf k}}).
\end{gather}

\subsubsection{\bf \textit{Spin-wave excitations using transverse magnetic susceptibility}}
The spin-wave excitations, which can be probed via the inelastic-neutron scattering (INS), can be obtained from the transverse-spin susceptibility ~\cite{dheeraj2, kovacic}
\bea
&&\chi_{\alpha \beta, \gamma \delta}(\q,\q^{\prime},i\omega_n)  \nonumber\\
&&=T \int^{\frac{1}{T}}_0{d\tau e^{i \omega_{n}\tau}\langle T_\tau
[{ S}^{+}_{\alpha \beta}(\q, \tau) {S}^{-}_{\gamma \delta} (-\q^{\prime}, 0)]\rangle}
\eea
defined for a multiorbital model. Further, it can be expressed in terms of Green's function as follows
\bea
&&\chi^{}_{\alpha \beta, \gamma \delta}(\q,\q,i\omega_n)  \nonumber\\
&&= \sum_{\k,i\omega^{\prime}_n} G^{0\uparrow}_{\alpha \gamma}
(\k+\q,i\omega^{\prime}_n+i\omega_n)G^{0\downarrow}_{\delta \beta} (\k,i\omega^{\prime}_n).
\eea
$\alpha$, $\beta$, $\gamma$, and $\delta$ are the orbital indices and take two values 1 and 2. The spin operator occurring in Eq. (14) is defined as
\be
{\cal S}^{i}_{\q}= \sum_{\bf k} \sum_{\sigma \sigma^{\prime}} \sum_{\mu \mu^{\prime}} d^{\dagger}_{\mu \sigma}(\k+\q)
E_{\mu \mu^{\prime}} \sigma^{i}_{\sigma \sigma^{\prime}} d_{\mu^{\prime} \sigma^{\prime}}(\k),
\ee
where $i = x,y,z$. $\hat{E}$ is a 2$\times$2 identity matrix in the orbital basis whereas $\sigma^i$s are the Pauli matrices for the spin basis. Because of the two orbitals, the transverse-spin susceptibility in the SDW state can be expressed in a matrix form
\be
\hat{{\chi}}^0(\q,i\omega_n) =
\begin{pmatrix}
 \hat{{\chi}}^0(\q,\q,i\omega_n) \,& \,\hat{{\chi}}^0(\q,\q+\Q,i\omega_n) \\
 \hat{{\chi}}^0(\q+\Q,\q,i\omega_n) \,& \,\hat{{\chi}}^0(\q+\Q,\q+\Q,i\omega_n)
\end{pmatrix},
\ee
where all the elements are themselves $n^2\times n^2$ matrices with $n = 2$. An element of the susceptibility matrix has contribution from several terms which include those arising because of Umklapp process and is given by
\bea
{\chi}^{0}_{\alpha \beta, \mu \nu} = \chi^{}_{\alpha \beta, \mu \nu}
+\chi^{}_{\bar{\alpha} \beta, \bar{\mu} \nu}+\chi^{}_{\alpha \bar{\beta}, \mu \bar{\nu}}
+\chi^{}_{\bar{\alpha} \bar{\beta}, \bar{\mu} \bar{\nu}}.
\eea
Here, $\bar{\alpha}$ stands for shift of momentum by $\Q$ for the orbital ${\alpha}$. The spin-wave excitations measured by the INS corresponds to the physical susceptibility defined by
\be
{\chi}^{psus}(\q, i\omega_n) = \sum_{\alpha \mu} {\chi}^0_{\alpha \alpha, \mu \mu} (\q, \q, i\omega_n).
\ee
The effect of on-site Coulomb interaction can be incorporated within the random-phase approximation (RPA) so that the susceptibility matrix is modified to
\be
\hat{{\chi}}_{\rm RPA}(\q, i \omega_n) = (\hat{{\bf 1}} - \hat{U}\hat{{\chi}}^0(\q,i \omega_n))^{-1} \hat{{\chi}}^0(\q,i \omega_n).
\ee
Here, $\hat{{\bf 1}}$ is a $2n^2 \times 2n^2$ identity matrix and the elements of block-diagonal matrix
$\hat{U}$ in the basis formed by ${\k}$ and $\k+\Q$ is
\begin{eqnarray}
&&{U}_{\alpha \beta; \gamma \delta}  \nonumber\\
&&= \left\{
\begin{array}{@{\,} l @{\,} c}
U \,\,& (\alpha=\beta=\delta=\gamma)\\
U-2J \,\, & (\alpha=\beta \ne \gamma=\delta)\\
J \,\,& (\alpha = \beta \ne \delta = \gamma)\\
J \,\, & (\alpha =\beta \ne \delta =\gamma)\\
0 \,\,& (\mathrm{otherwise})
\end{array}\right..  \nonumber\\
\end{eqnarray}
Analytic continuation $i \omega_n \rightarrow \omega + i \eta$ is performed with $\eta$ as 0.05.

\section{Results}
\begin{figure} [hb]
    \centering
     \includegraphics[scale = 1.0, width = 8.2cm]{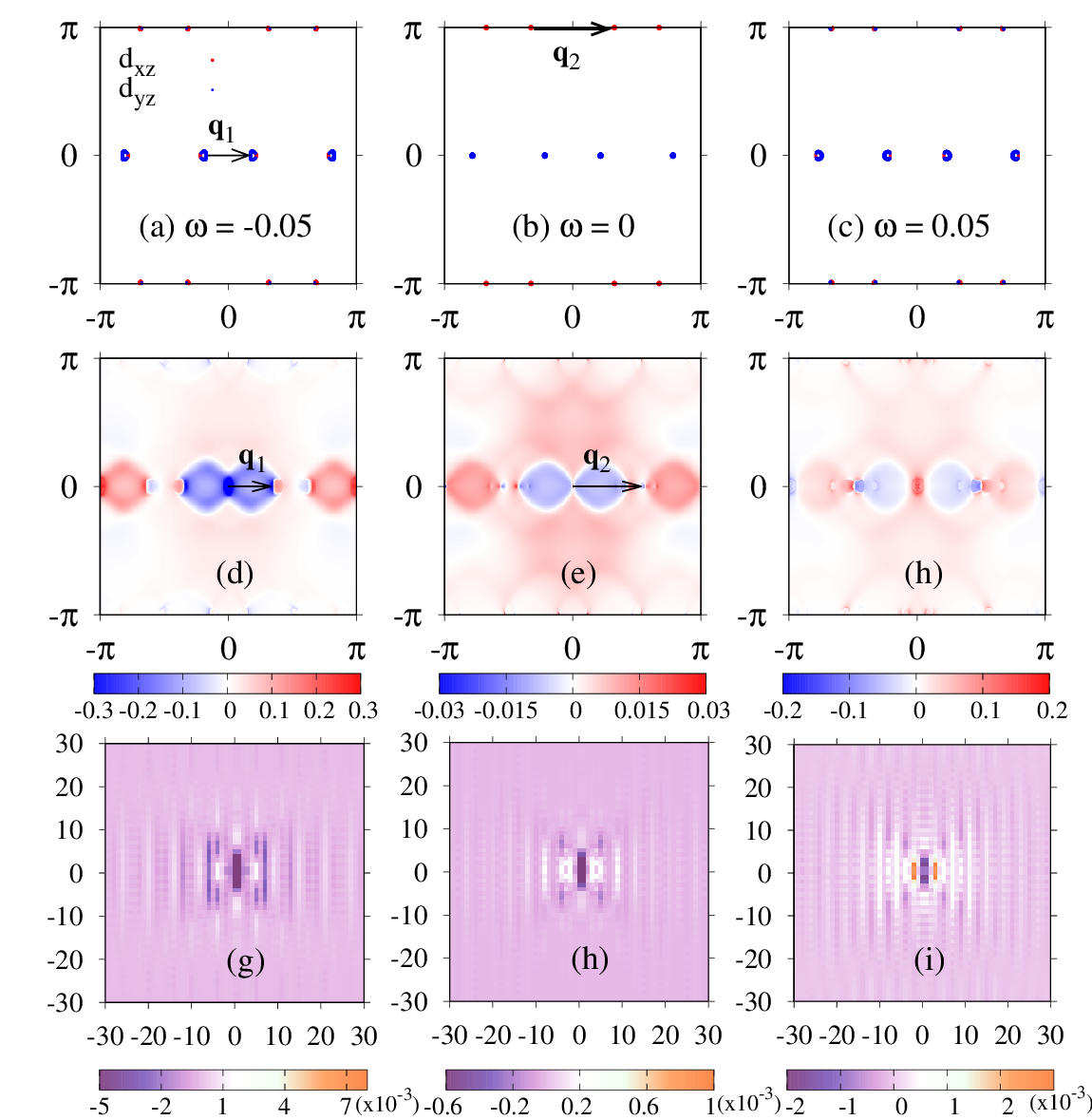}
    \caption{Constant energy contours (CECs) of the spectral function $\mathcal{A}({\k}, \omega)$ plotted for energies (a) $\omega$ = -0.05, (b) $\omega$ = 0, and (c) $\omega$ = 0.05 in the semimetallic SDW state. ${\q}_1$ and ${\q}_2$ refers to the scattering vectors corresponding to intraorbital scattering for two pairs of Dirac cones located along $k_y = 0$ and $k_y = \pi$, respectively. Momentum-space and real-space QPI maps are shown in the second and third rows respectively.}
    \label{1}
\end{figure}

\begin{figure} [hb]
    \centering
    \includegraphics[scale = 1.0, width=0.95\linewidth]{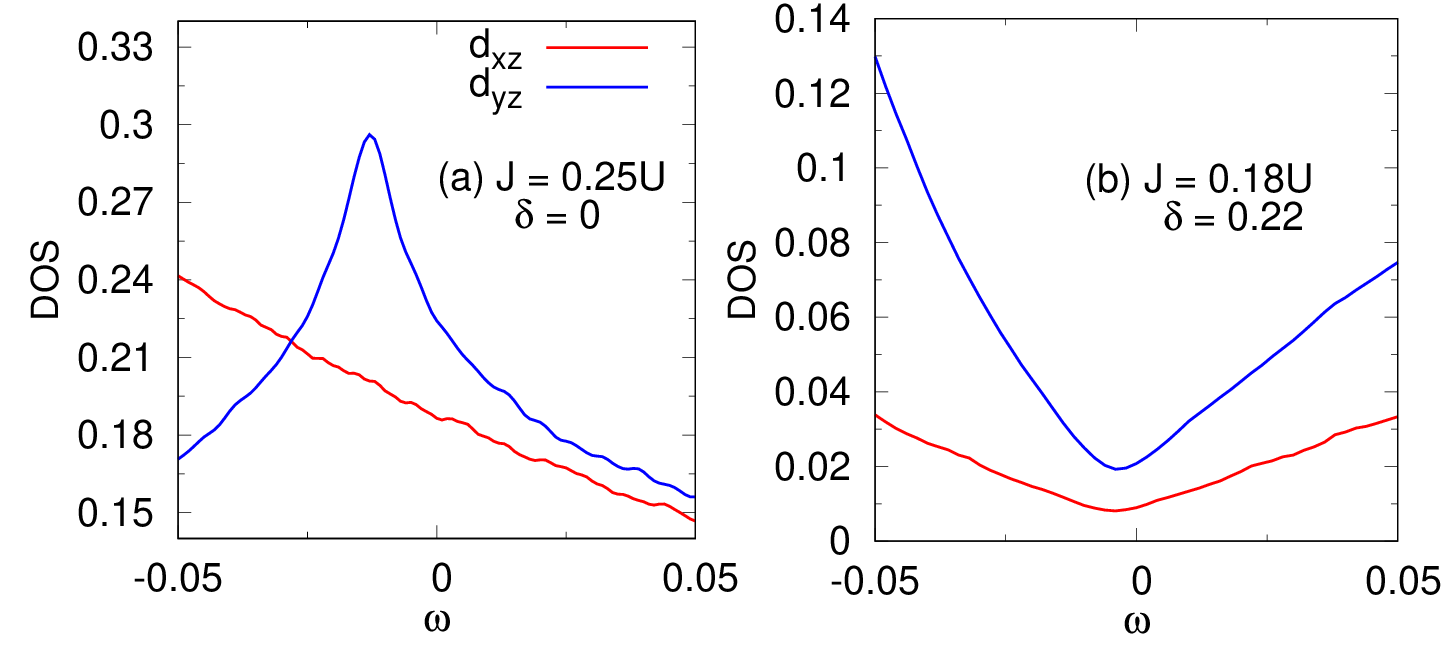}
    \caption{Orbital-resolved DOS as a function of energy $\omega$ for $U = 4.0$ in the (a) ordinary metallic state for $\delta = 0$ and (b) Dirac semimetallic state for $\delta = 0.22$.}
    \label{2}
\end{figure}

Fig.~\ref{1} shows constant-energy contours, QPI patterns as well as modulations in the local density of states (LDOS) for an energy range spanning from $\omega$ = -0.05 to 0.05  with a step size of 0.05. The interaction parameters are chosen to be $U = 4.0$ and $J = 0.18U$ so as to obtain a Dirac semimetallic state in a self-consistent manner, which is possible near $\delta \approx 0.22$. In the Dirac semimetallic state, there are no additional bands, which cross the Fermi level unlike in the ordinary metallic SDW state. This makes the analysis possible for the contribution to the anisotropy that originates purely from the Dirac cone.

\begin{figure} [t]
    \centering
    \includegraphics[width=0.95\linewidth]{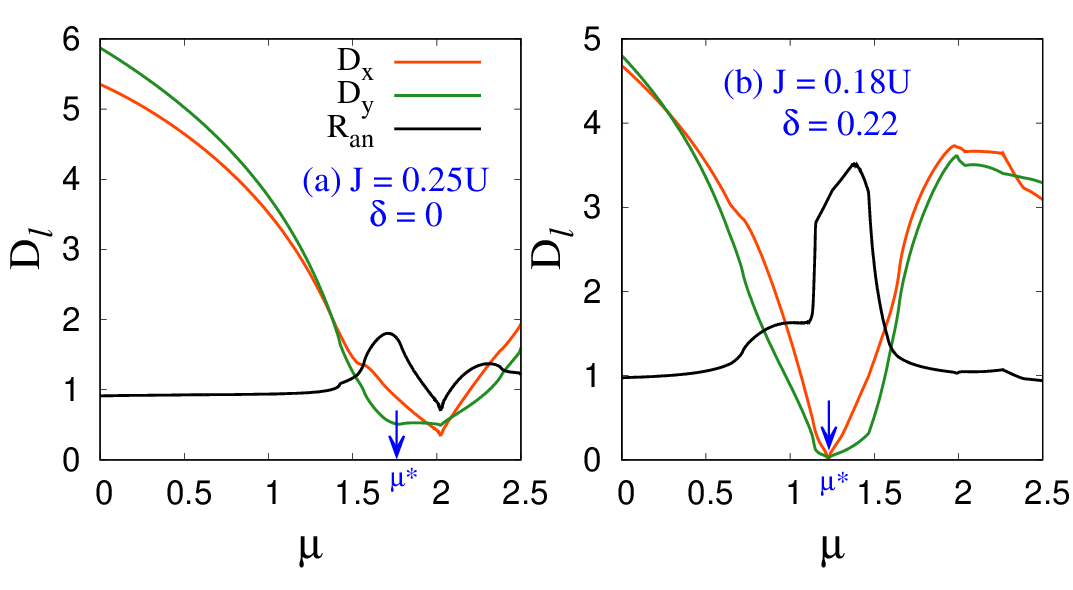}
    \caption{Drude weights along \textit{l} = $x$- and $y$-directions with anisotropy parameter defined as $R_{an} = \frac{D_x}{D_y}$ in the (a) ordinary metallic and (b) Dirac semimetallic state.}
    \label{3}
\end{figure}

\begin{figure} [b]
    \centering
    \includegraphics[width=0.95\linewidth]{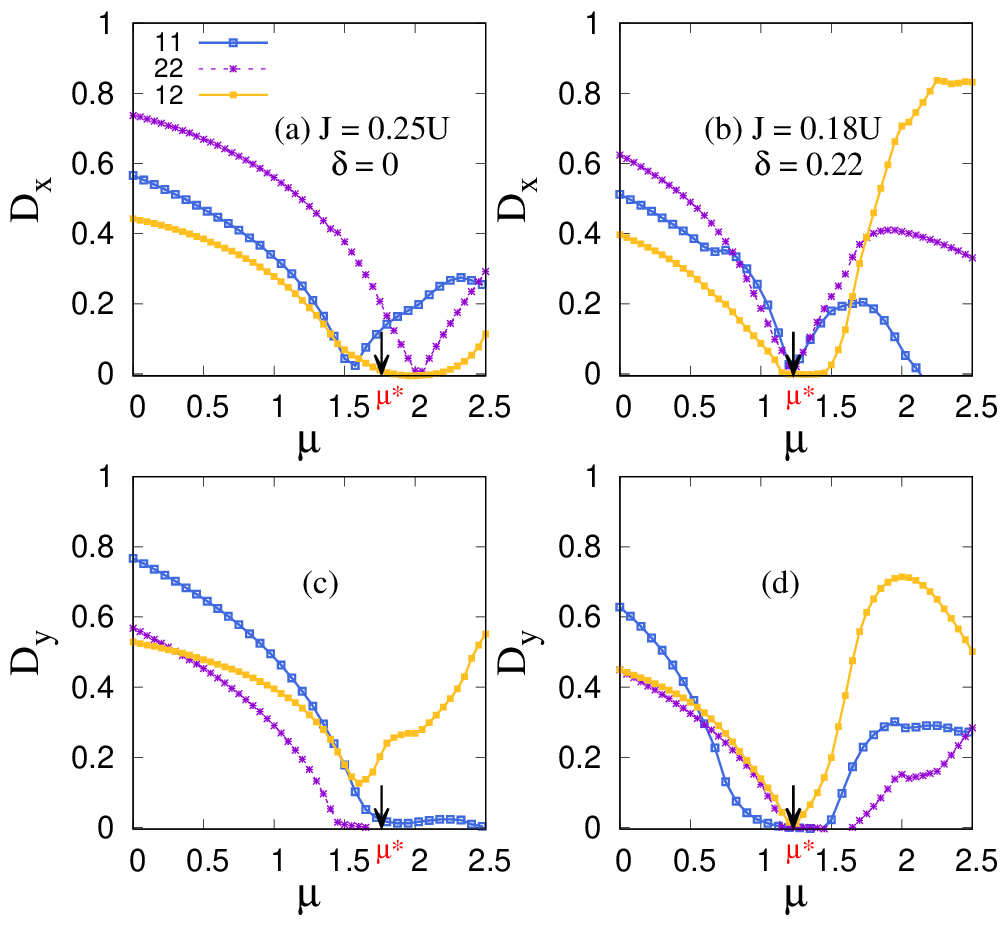}
    \caption{Various components of the Drude weights along the $x$-direction for ordinarly metallic state ((a) and (c)) and along the $y$-direction for the Dirac semimetallic states ((b) and (d)).}
    \label{4}
\end{figure}

The first row shows the CECs with orbital distributions indicated by different colors. The cross sections of the Dirac cones appear to be of a single color in the vicinity of $\omega = 0$ though they are not. They appear so because of the non-circular shape of the pockets, which is not clearly visible because of small size. Moreover, the weight of the two orbitals is also not equal for a given pair of Dirac cones, particularly when the orbital splitting is incorporated into the tight-binding  part of the model in order to obtain the Dirac semimetallic SDW state (Fig.~2). More clearly seen for higher $\omega$, the pockets along $k_y$ = 0 are primarily dominated by $d_{yz}$ orbital, with a smaller contribution from $d_{xz}$. The two pairs of Dirac cones, one along $k_y = 0$ and the other along $k_y = \pi$, have opposite dominating orbitals. Therefore, it is expected that the inter-pair scattering of the Dirac cone is suppressed, which would otherwise have led to a modulation along the diagonal direction. Incidentally, there exists only a pair of Dirac cones in the SDW state obtained within most of the five-orbital models when a  realistic-interaction parameter regime is chosen, and therefore such complications do not arise there.

\begin{figure}[hb]
    \centering
    \includegraphics[width=0.95\linewidth]{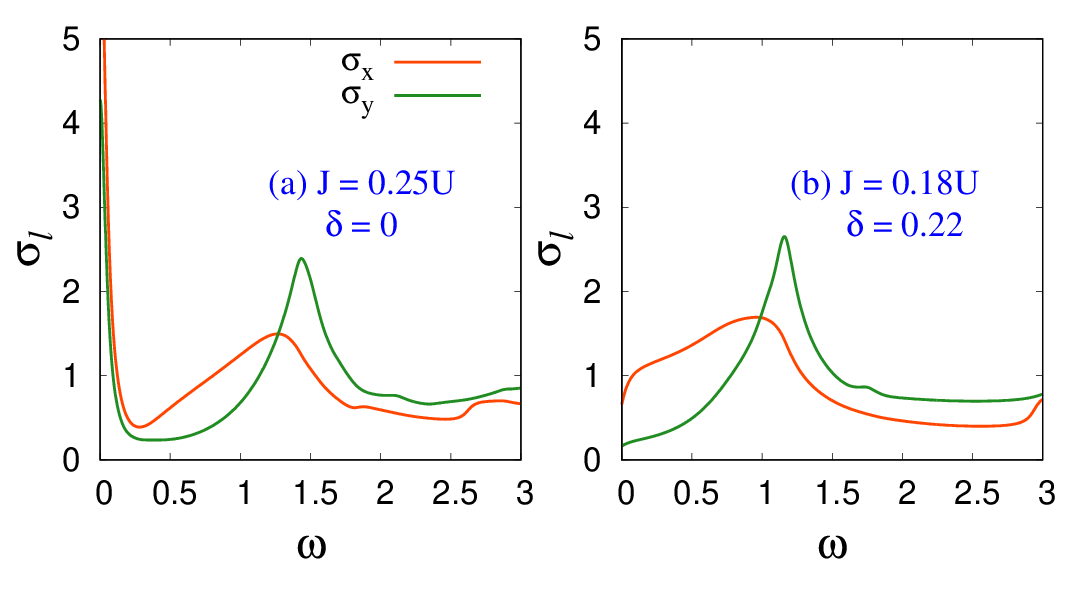}
    \caption{Conductivity along x- and y-direction in the (a) metallic and (b) semimetallic SDW state. }
    \label{5}
\end{figure}

The dominance of the pair of Dirac cones along $k_y = 0$ is clearly visible in the QPI patterns (Fig.~\ref{1}(d-f)). There exist strong modulating patterns along $q_y = 0$ arising as a result of the scattering vector ${\q}_1$ between the pair of the Dirac cones along $q_y = 0$. There is another scattering vector $\q_2$, which may also be expected to contribute to the patterns as a result of the scattering between another pair of Dirac cones lying along $k_y = \pi$. However, $\q_1$ shows its dominance in the entire energy range considered, which is also reflected in the real space QPI patterns (Fig. \ref{1} (g-i)), where the periodicity is determined solely by ${\q}_1$ scattering. This behavior can be understood in the context of orbital-resolved DOS as a function of energy (Fig. \ref{2}(b)). Notably, the contribution of $d_{yz}$ orbital exceeds in comparison to the $d_{xz}$ orbital in the whole considered energy regime (Fig.~2). As a result, the scattering between pockets with a dominant $d_{yz}$ orbital character governs the overall QPI pattern.

At negative energy ($\omega = -0.05$), the pockets along $k_y = 0$ with a dominant $d_{yz}$ contribution are separated by a distance of $\sim \pi/3$, resulting in a periodic modulation of $\sim6a$ (Fig. \ref{1} (g)). For the other two cases, the separation of the pockets along $k_y = 0$ change slightly to $  \sim \pi/2$, producing a modulation with a periodicity $\sim$ 4$a$ (Figs. \ref{1} (h), (i)). For the Dirac cone located along $k_y = 0$, the difference in the magnitude of the scattering vectors for positive and negative $\omega$ arises because the band dominated by $d_{xz}$ orbital is approximately given by $k_x \sim c$ in the vicinity of the Dirac points whereas the other band dominated by the $d_{yz}$ orbital has a slope $\sim \pi/3$~\cite{garima}. Thus, the $d_{yz}$ dominated regions shift across the Dirac point as the energy changes from negative to positive.

\begin{figure}[t]
    \centering
    \includegraphics[width=0.98\linewidth]{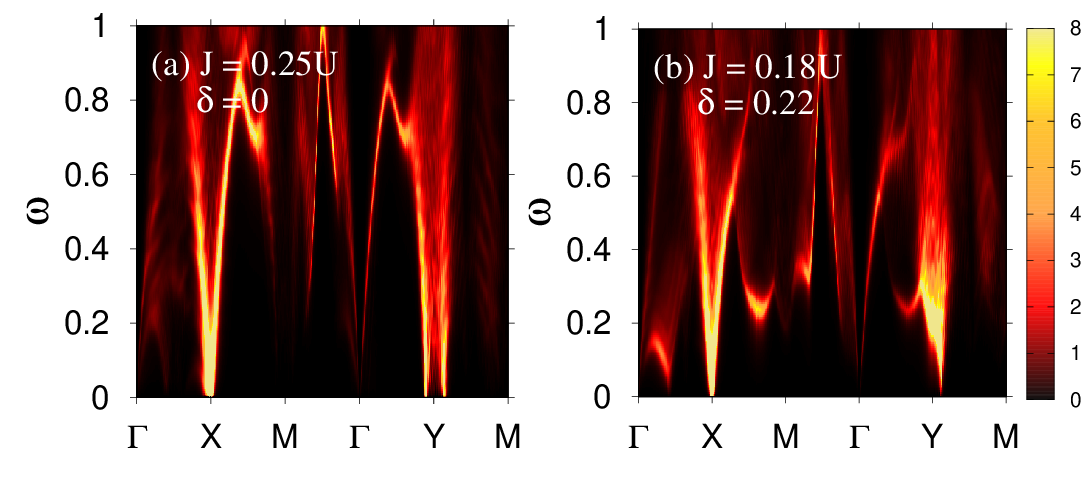}
    \caption{Spin-wave excitations along high-symmetry directions for $U = 4.0$ in the (a) metallic state when $J = 0.25U$ and (b) semimetallic state when $J = 0.18U$.}
    \label{6}
\end{figure}

\begin{figure}[t]
    \centering
    \includegraphics[width=1.0\linewidth]{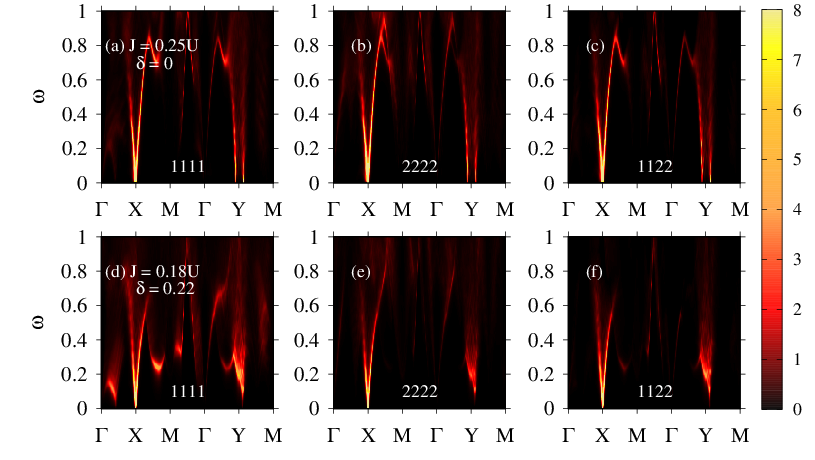}
    \caption{Various components of spin-wave excitations along the high-symmetry directions. The upper row demonstrates components in the metallic state while the bottom row in the Dirac semimetallic state.}
    \label{7}
\end{figure}

Next, we discuss the role of the Dirac cone in the contribution to the anisotropy of optical conductivity. Fig.~\ref{3} shows the Drude weight along the $x$-direction with the magnetic moments ordered antiferromagnetically as well as along $y$-direction with magnetic moments ordered ferromagnetically. Fig.~\ref{3}(a) shows the Drude weight when the SDW state is an ordinary metal while Fig.~\ref{3}(b) shows the Drude weight when the SDW state is a Dirac semimetal. We find that in the ordinary metallic SDW state, the anisotropy parameter defined as a ratio of two Drude weights is nearly one, away from the chemical potential $\mu^*$ corresponding to the band filling of $n = 2$. $\mu^*$ is obtained through the self-consistent process whereas variation of the chemical potential is only for the illustration purpose where the self-consistently obtained parameters are fixed to be that corresponding to $\mu^*$.

Notably, the band filling $n = 2$ in the unordered state of the two-orbital model reproduces the Fermi surfaces obtained via bandstructure calculations except for an extra pocket near ($\pi, \pi$) and significant contribution of $d_{xy}$ orbital at the Fermi level.
The anisotropy parameter deviates from unity only in the immediate vicinity of $\mu^*$. More importantly, the Dirac points are also in the vicinity of the Fermi level corresponding to this band filling though on the opposite of the Fermi level. In other words, if the Dirac points were absent the anisotropy in the Drude weight would have been weak. This can be seen in Fig.~3(b), where the Drude weight along different directions are plotted for the Dirac semimetallic SDW state with the Dirac points located at the Fermi level. In this case, the Drude-weight anisotropy maximizes in the vicinity of $\mu^{*}$.

A much clearer insight into the origin of anisotropy can be obtained via Fig.~\ref{4}, which shows individual components of the anisotropy such as $D^{11}_x$, $ D^{22}_x$, etc. The superscripts 1 and 2 correspond to $d_{xz}$ and $d_{yz}$ orbitals, respectively. Along the $x$ direction, in both $D^{11}_x$ and $ D^{22}_x$, there is a linear drop as one approaches $\mu^{*}$ and a rise thereafter. On the other hand, both get flattened near $\mu^{*}$ along $y$. An opposite trend is shown by $D^{12}$, however, the two contributions, each from  $D^{11}_x$ and $ D^{22}_x$ results in the Drude weight enhanced along $x$-direction as compared to $y$-direction. As expected, the difference in the optical conductivity along the two orthogonal directions is more prominent in the low-energy region (see Fig.~\ref{5}). Moreover, this difference is further enhanced in the Dirac semimetallic SDW state. However, one interesting point to be noted is that, near $\omega \sim 1$ and beyond, $\sigma_y$ becomes larger than $\sigma_x$ as one would have expected conventionally that the conductivity should be better along the ferromagnetic direction.

Next, we examine the difference in the spin-wave excitations in the ordinary metallic and Dirac semimetallic SDW states. Our finding suggests that there is no significant difference in the nature of spin-wave excitations in these two states except that the damping of high-energy spin-wave excitations is more prevalent in the latter. Moreover, we may also notice that the spectral weight appears to be transferred to the low-energy region. This peculiar behavior can be understood if we look at various elements of magnetic susceptibility such as $\chi_{11,11}$, $\chi_{22,22}$, $\chi_{11,22}$, etc. The transfer of spectral weight to the low-energy region and damping is more prominent in the component $\chi_{11,11}$ instead of $\chi_{22,22}$. The behavior can be attributed to the orbital-splitting parameter $\delta$ because of which the occupancy of $d_{xz}$ orbital is more than half filling whereas that of $d_{yz}$ orbital is less than half filling. We find the magnetization in both the orbitals $\sim$ 0.22, which means that the relative polarization $(n_{\uparrow} - n_{\downarrow})/(n_{\uparrow} + n_{\downarrow})$ is larger for the $d_{yz}$ orbital intead of the $d_{yz}$ orbital. The same phenomenon may be observed in the five-orbital model where a major contribution to the spin-wave damping may occur because of the orbitals which has band filling more than half filling. For the undoped case and a realistic regime of interaction parameters, it has been noted earlier that $n_{xy}$ is close to half filling whereas $n_{xz}$ deviates more from half filling in comparison to $n_{yz}$ orbital though both being slightly larger than 1. Thus, the intra-orbital component of susceptibility corresponding to the $d_{xz}$ orbital may contribute majorly in the damping of spin-wave excitations just like the two-orbital model considered here.

\section{ Summary and conclusions}
ARPES measurements in the SDW state of iron pnictides have revealed that the Dirac cones or the Dirac points may not be far away from the Fermi surface~\cite{richard,watson}. To the extent, they can affect the electronic properties is not clearly established. Through the current work, we have attempted to unravel the role of the Dirac cones by using a minimal two-orbital model. In this model, the Dirac points, with the help of orbital splitting, can be brought at the Fermi level while there is no other band crossing, an ideal scenario that can provide insight into the role of Dirac cone in the anisotropic properties. On the other hand, achieving the same in the five-orbital model is a very challenging task mainly because of a larger hole pocket around $\Gamma$.

Previously, the origin of the anisotropy in the QPI patterns was attributed to the reconstruction of the Fermi surfaces as a result of the formation of the SDW state. However, the Fermi-surface reconstruction is also accompanied by redistribution of orbital weights, and therefore the assumption that the nature of the impurity potential is such that it preserves the orbital state of the quasiparticle plays a key role in resulting QPI patterns. It may be noted that despite different orbitals dominating different regions of the pair of cones along $k_y = 0$ or $k_y = \pm \pi$, the $d_{yz}$ orbital is the dominant one for the pair along $k_y = 0$, whereas the cones located along the $k_x =  \pm \pi $ are dominated by the $d_{xz}$ orbital. Therefore, the orbital-state preserving requirement dictates that the scattering along $k_x$ should be the dominant one. Although illustrated in the minimal two orbital model, we believe that the origin of highly-anisotropic QPI patterns in the SDW state of iron pnictides can be explained provided that the Fermi surfaces consist of only sections of Dirac cones, which is not in disagreement with ARPES measurements. At the same time, there should not be any other large pockets. Even if there are such pockets, their spectral weight should be significantly suppressed so that the spectral weight is very small as compared to that of the pockets associated with the Dirac cones. Presence of large pockets with significant spectral weight is expected to weaken or even destroy the one-dimensional character of QPI patterns. Therefore, any tight-binding model should have aforementioned features of the Fermi surfaces in the SDW state obtained with a realistic set of interaction parameters in order to explain the extent of anisotropy in QPI patterns as well as other electronic properties.

Unlike $d_{xz}$ and $d_{yz}$ orbitals, the sections of the Dirac cone not far away from the Dirac point are dominated by $d_{yz}$ and $d_{xy}$ orbitals within a five-orbital model. Moreover, the pair of cones exists only along $k_y = 0$. Thus, a one-dimensional modulation in the quasiparticle interference similar to what is obtained in the two-orbital model can be realized easily provided that the pocket around $\Gamma$ is absent. We also note that unlike the QPI where the one-dimensional patterns are either weakened or destroyed by the larger pocket around $\Gamma$, the anisotropy of Drude weight is unlikely to be affected by the larger pocket around $\Gamma$ in the five-orbital model because of the overall dominance by $d_{xz}$ orbital. Whatever anisotropy exists, it is a consequence of the elliptical shape of the pocket. On the other hand, the Dirac cone dominated either by $d_{yz}$ or $d_{xy}$ orbitals contributes the most to the anisotropy.

To conclude, we used a minimal two-orbital model to demonstrate the contribution from the Dirac nodes to the anisotropic electronic properties of the iron pnictides in the SDW state. We find that the highly-anisotropic distribution of the orbital weights along the Dirac cone may be a significant contributor to the unusually large anisotropy in the optical conductivity. As another important consequence of Dirac cones not being far away from the Fermi surface, we find nearly one-dimensional modulation for the quasiparticle interference though with modulation wavevector $\sim \pi/3-\pi/2$ in the momentum space and nearly four to six times the interatomic distance between two nearest neighbor iron atoms. Presence of Dirac points was found not to introduce any unusual behavior in the low-energy spin-wave excitations, which may be significantly different from the ordinary-metallic state.

\section*{ACKNOWLEDGEMENTS} 
D.K.S. was supported through DST/NSM/R\&D\_HPC\_Applications/2021/14 funded by DST-NSM and start-up research grant SRG/2020/002144 funded by DST-SERB.

\end{document}